\documentclass[pre,aps,twocolumn]{revtex4}
\usepackage{graphicx}

\def\pt{$p_T$ }
\def\aa{Au+Au }
\def\da{$d$+Au }

\begin{document}
\title{\bf Proton Production in \da Collisions and
the Cronin Effect}
\author{Rudolph C. Hwa$^1$ and  C.\ B.\ Yang$^{1,2}$}
\affiliation{$^1$Institute of Theoretical Science and Department of
Physics\\ University of Oregon, Eugene, OR 97403-5203, USA}
\affiliation{$^2$Institute of Particle Physics, Hua-Zhong Normal University,
Wuhan 430079, P.\ R.\ China}

\begin{abstract}
Proton production in the intermediate \pt region in
\da collisions is studied in the parton recombination model. The
recombination of soft and shower partons is shown to be important in
central collisions, but negligible in peripheral collisions. It is
found that the large nuclear modification factor for proton
production can be well reproduced by a calculation of the 3-quark
recombination process.
\end{abstract}
\maketitle

In a previous paper \cite{hy} we have shown that the
Cronin effect
\cite{jc} on pion production in \da collisions can be understood in
terms of the recombination of the soft and shower partons without \pt
broadening by multiple scattering in the initial state.  In this
short note we extend the consideration to proton production and show
that the same effect can similarly be interpreted.

The Cronin effect can best be displayed by the nuclear modification factor
$R_{CP}(p_T)$ that is the ratio of central-to-peripheral yields
appropriately scaled by the average number of binary collisions
$\left<N_{\rm Coll}\right>$.  It is found in the PHENIX experiment that
$R^p_{CP}$ for proton reaches a value roughly $2$ for $2<p_T<3$
GeV/c, even higher than that for pion at $\sim 1.4$ \cite{fm}.  Such a
behavior of the enhancement effect is hard to interpret, if hadrons
produced at intermediate $p_T$ are the consequences of fragmentation
of hard partons produced at higher $p_T$.  Indeed, since there is no
energy loss in \da collision, one would expect $R_{CP}\sim 1$ for
both pion and proton on the grounds that fragmentation outside the
cold medium should be independent of the impact parameter.  Thus
the observed $R^{\pi, p}_{CP}$ strongly suggests the dependence of
the hadronization mechanism on the medium.  In our view recombination
is that mechanism, which, on the one hand, provides a way to describe
fragmentation in terms of shower partons \cite{hy2}, and, on the
other, can take into account the coalescence of soft and shower
partons to form hadrons in the intermediate $p_T$ range \cite{hy3}.

For the proton spectrum at low $p_T$, one should be careful about the
mass effect. The low-$p_T$ region is, however, not the main part of
our
work where the model has any predictive power.  As in \cite{hy,hy3},
we fit the data in that region and use our model to predict the
hadronic spectra at intermediate and high $p_T$.  We have found that
for the purpose of data fitting at low $p_T$ our 1D formulation of
the recombination process is quite adequate when only the kinematical
variables are suitably modified to account for the proton mass.
Thus we start with the invariant inclusive distribution for proton
formation at midrapidity in the recombination model \cite{hy3,rh}
\begin{eqnarray}
p^0{dN_p  \over  dp} = \int {dp_1 \over
p_1}{dp_2 \over p_2} F (p_1, p_2, p_3) R_p(p_1, p_2, p_3, p)  ,
\label{1}
\end{eqnarray}
where all momentum variables $p_i$ and $p$ are transverse
momenta, and $p^0$ denotes the energy of the proton.  Since the
parton masses are set to zero, we continue to use $p_i$ for their
energies. $F (p_1, p_2, p_3)$ is the joint distribution of $u, u,$ and
$d$ quarks at
$p_1, p_2$ and $p_3$, respectively.  $R_p(p_1, p_2, p_3, p)$ is the
recombination function  for a proton with momentum $p$ \cite{hy3,hy4}
\begin{eqnarray}
R_p(p_1, p_2, p_3, p) = g (y_1y_2)^{\alpha + 1}y_3^{\gamma +1}
\delta \left(\sum_i y_i - 1\right) \quad ,
\label{2}
\end{eqnarray}
where $y_i = p_i/p$, $\alpha = 1.75$, $\gamma = 1.05$, and
\begin{eqnarray}
g = \left[6 B (\alpha + 1, \alpha + \gamma +2) B (\alpha + 1,\gamma
+1) \right]^{-1} \ ,
\label{3}
\end{eqnarray}
$B(a, b)$ being the beta function. As always in the recombination
model, the main issue is about the distribution of the quarks that recombine.
Here, it is $F (p_1, p_2, p_3)$.

Following the same notation used in \cite{hy3} for \aa  collisions,
we write schematically
\begin{eqnarray}
F = {\cal TTT} + {\cal TTS} + {\cal TSS}  +
{\cal SSS} ,
\label{4}
\end{eqnarray}
where all the shower partons $\cal{S}$ are from one hard parton jet.
Shower partons from different jets are ignored here for RHIC energies.
$\cal{ T}$ denotes thermal parton, even though in \da  collisions the
notion of thermal equilibrium may not be justified.  To preserve the
same notation as in \cite{hy3}, we continue to use $\cal{ T}$ to signify
the soft partons that are not associated with the shower components of
a hard parton and are loosely referred to as thermal partons when
convenient.  The ${\cal SSS}$ term in Eq.\ (\ref{4}), when convoluted
with $R_p$ in Eq.\ (\ref{1}), gives rise to what is usually regarded as
the fragmentation of a hard parton into a proton \cite{hy2}.  The
${\cal TTT}$ term comes entirely from the soft partons, while ${\cal
TTS}$ and ${\cal TSS}$ accounts for the interplay between the
thermal (or soft) and shower partons.

As in \cite{hy,hy3}, we can calculate the distributions of the shower
partons ${\cal S}$ from the QCD processes of producing hard partons
and their induced shower partons.  We cannot calculate the thermal
component $\cal{ T}$, which is deduced from fitting the low-$p_T$
data.  Thus what we can calculate that is new is only the effect of the
${\cal TTS} + {\cal TSS}$ terms at intermediate $p_T$, knowing that the
${\cal SSS}$ term dominates at very high $p_T$.

For $\cal{ T}$ we use the same parametrization as before
\cite{hy,hy3}, and write
\begin{eqnarray}
{\cal T}(p_1) = p_1{dN^{{\cal T}}_q  \over  dp_1}
= Cp_1 \exp (-p_1/T),
\label{5}
\end{eqnarray}
where $T$ should be regarded as just an inverse slope.  The
thermal contribution to the proton spectrum arising from ${\cal
TTT}$ recombination is then
\begin{eqnarray}
{dN^{\rm th}_{\pi}  \over  pdp} =  {C^3 \over
6} {p^2 \over p^0}  e^{-p/T} { B (\alpha + 2,  \gamma
+2) B (\alpha + 2,\alpha + \gamma +4) \over B (\alpha + 1,
\gamma +1) B (\alpha + 1,\alpha + \gamma +2) } ,
\label{6}
\end{eqnarray}
which differs from a similar formula in \cite{hy3} by only the presence
of $p^0$, instead of $p$.  For the other three terms in Eq.\
(\ref{4}) that involve ${\cal S}$, the contributions to the proton
spectrum are the same as those in \cite{hy3} except that each
equation should be multiplied by the factor $p/p^0$ on the RHS
and the factor $\xi$ should be omitted.  The latter refers to the
mean energy loss in \aa collisions, and should be absent in
\da collisions. Also the hard parton distributions $f_i(k)$ in
\cite{hy3} should be changed to the corresponding ones for \da
collisions, as given in \cite{hy}.  What is to be emphasized is that
there
are no free parameters to adjust in those terms.  The shower parton
distributions are the main input, and they have previously been
determined in \cite{hy2}.

We must now determine $C$ and $T$ by fitting the low-$p_T$ data
using Eq.\ (\ref{6}) for both central and peripheral collisions.  The data
available are from PHENIX, given as figures online \cite{fm2}.  We fit
them in the region $0.5 < p_T < 1.5$ GeV/c and the results are shown
by the light solid lines in Fig.\ 1 (a) and (b) for 0-20\% and 60-90\%
centralities, respectively.  The values determined are $C = 11.5\ (8.0)$
(GeV/c)$^{-1}$ and $T = 0.24\ (0.21)$ GeV/c for 0-20\% (60-90\%)
centrality.  The contributions from ${\cal TTS} + {\cal TSS}$ and
${\cal SSS}$ components are determined without free parameters, and
are shown by the dashed and dash-dot lines in the same figures.  The
sum of all four components are shown by the thick solid lines.
Evidently, they agree with the data very well.

\begin{figure}[tbph]
\includegraphics[width=0.45\textwidth]{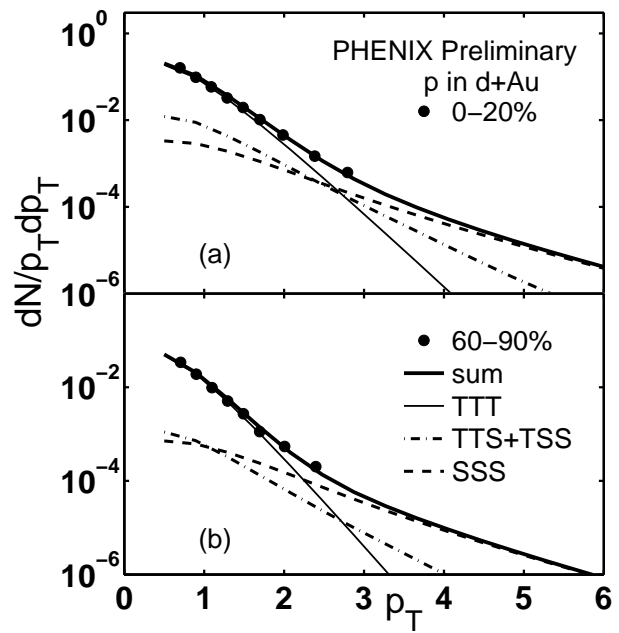}
\caption{(a) Proton transverse momentum distribution in \da
collisions at 0-20\% centrality. The preliminary data are from
\cite{fm2}.
(b) Same as in (a), but for 60-90\% centrality.}
\end{figure}

We note that at 0-20\% centrality the thermal-shower  $({\cal TTS} +
{\cal TSS})$ contribution crosses over the fragmentation $({\cal SSS})$
   component at $p_T \approx 2.5$ GeV/c, roughly the same as in the
case of pion production in \da collisions \cite{hy}.  However,  the
thermal ${\cal TTT}$ contribution is roughly the same as each of the
$({\cal TTS}+{\cal TSS})$ and $\cal SSS$ contributions  at the
cross-over point, whereas for pion production the thermal ${\cal TT}$
contribution is much lower than $\cal TS$ and $\cal SS$ at the same
point.  Thus the thermal
contribution to proton formation dominates over a wider range of
$p_T$ than that for pion.  That is because of the $C^3$ dependence
in Eq.\ (\ref{6}), and is consistent with the findings in Refs.\ \cite{gr,fr},
where the recombination of thermal partons can account for the large
$p/\pi$ ratio up to  $p_T \approx$ 3-4 GeV/c \aa collisions.
The same cannot be said about 60-90\% centrality in \da
collisions.  The cross-over between  ${\cal TTS} + {\cal TSS}$ and
${\cal SSS}$ occurs at $p_T \approx 1$ GeV/c, where the
distributions are far lower than the thermal contribution.  Throughout
all $p_T$ in Fig.\ 1 (b) the $({\cal TTS} + {\cal TSS})$ component is
negligible, a
situation that is similar to what we have found separately for either
$p$ or $\pi$ production in $pp$ collisions.  The reason is
the  substantial reduction of the soft parton environment in peripheral
\da or $pp$ collisions so that the $({\cal TTS+TSS})$ component  does
not develop any strength relative to the $\cal SSS$ component. We
should, however, note parenthetically that the low density of soft
partons that are produced in $pp$ collisions is sufficient to make
the jet structure different from that in $e^+e^-$ collisions, which
have no soft partons of the same origin at all.

The fact that our calculated results agree well with the data for both
0-20\% and 60-90\% centralities implies that the ratio
$R^p_{CP}(p_T)$, defined by
\begin{eqnarray}
R^p_{CP}(p_T) = {\left<N_{\rm Coll}\right>_{60-90\%}  dN_p/dp_T
(0-20\%) \over
\left<N_{\rm Coll}\right>_{0-20\%} dN_p/dp_T
(60-90\%)}
\label{7}
\end{eqnarray}
must also agree with the data.  That agreement is shown explicitly in
Fig.\ 2 by the solid line, where the data are from Ref.\ \cite{fm}.
The calculated curve approaches 1 at high
$p_T$, where the yields are dominated by the fragmentation of hard
partons $({\cal SSS})$, which is independent of the soft partons.  The
dashed line in Fig.\ 2 shows the contribution to $R^p_{CP}$ from the
thermal components only.  The small difference in $T$ for the two
centralities results in an exponential growth in $p_T$ as can be seen
directly from Eq.\ (\ref{6}).  Thus the effect of thermal-shower
recombination is the damping of that exponential increase in
$R^p_{CP}(p_T)$, as shown by the solid line.  Because of the
ineffectiveness of the thermal-shower contribution at 60-90\%
centrality, that damping does not take place until $p_T$ reaches near
2 GeV/c.  By then $R^p_{CP}$ is already greater than 1.7, which is
higher than $R^{\pi}_{CP}$.  Thus the origin of $R^p_{CP}$ being
greater than $R^{\pi}_{CP}$ is mainly in the ${\cal TTT}$
recombination for proton being more sensitive to centrality than
${\cal TT}$ for pion.  The role of shower partons is limited in that
comparison.

It should be pointed out that our fitting procedure in the
determination of $C$ and $T$ has not ignored $R^p_{CP}$ as an
outcome.  Since Fig.\ 1   has logarithmic vertical scale, those
data points can determine $C$ and $T$ only within narrow ranges.
The data in Fig.\ 2 involve their ratios and are in linear scale.  Thus
the parameters can be more accurately determined by including Fig.\
2 in the fit.  Since the distributions of the soft  partons
at low $p_T$ are not generated from first-principle calculations, we
have taken the liberty to use all the data available to make the best
determination of them.  Our predictions are only
for the behavior at \pt above the region that is dominated by the soft partons.

The values of $C$ and $T$ obtained here should be compared with
those determined from the pion spectrum \cite{hy}.  They are
$C_{\pi} = 12\ (5.65)$ (GeV/c)$^{-1}$ and
$T_{\pi} = 0.21\ (0.21)$ GeV/c for 0-20\% (60-90\%) centrality,
where the subscripts $\pi$ are added for distinction.
Whereas there
is no essential dependence of some of those parameters on whether the
formed hadrons are pions or protons, i.e., $C_p\simeq C_{\pi}$ at
0-20\% and $T_p=T_{\pi}$ at 60-90\%, there exist some
significant differences in others:  $C_p = 8.0$, $C_{\pi} =5.65$ at
60-90\% and $T_p = 0.24$, $T_{\pi} =0.21$ at 0-20\%.  The species
dependence of those parameters reflects the general properties of the
spectra, especially at $p_T < 1$ GeV/c.  In \cite{ssa} it is shown in
Au+Au collisions that the low-$p_T$ spectra can be fitted by
exponential behavior in $m_T$ with the inverse slope increasing
linearly with hadronic mass at a rate that increases with centrality.  It
strongly suggests hydrodynamical expansion radially, which is well
known to exist.  It means that a significant portion of the
hydrodynamical fluid is hadronic at very low $p_T$.  Only the
portion that remains partonic at  $p_T$ around 1 GeV/c and above
are available for recombination with the shower partons.  How much
of this scenario is valid  (and  in what quantitative way)  for \da
collisions is not known.  Since our hadronization model does not
treat the hydrodynamical expansion phase, we can only take the
thermal hadrons in the $0.5 < p_T < 1.5$ GeV/c range as observed,
but not lower, to determine our parameters for the thermal partons
${\cal T}$.  The species dependence of some of the $C$ and $T$
parameters is clearly a result
of our inadequacy in subtracting out the already-formed hadrons
from the medium at low $p_T$, since the thermal partons ${\cal T}$
should have no knowledge of what hadrons they are to form.    Given
our inability to treat very low $p_T$ physics, we can only regard
what we have done as a determination of the intermediate $p_T$
behavior in the separate cases of specific hadrons  without a way to
enlighten the problem of overall species dependence at very low $p_T$
in a broader scheme.

\begin{figure}
\includegraphics[width=0.45\textwidth]{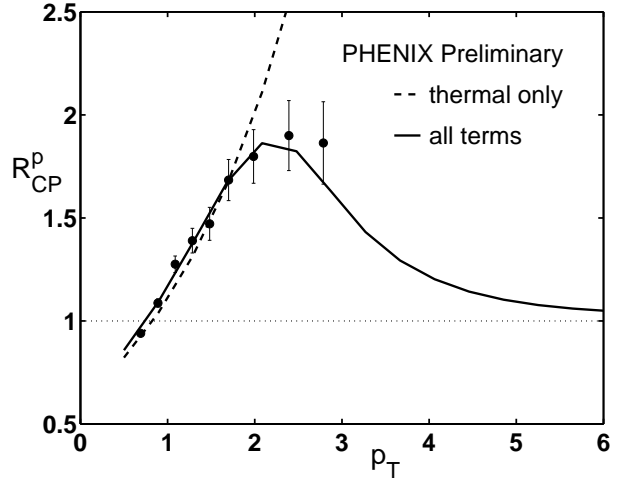}
\caption{
 Nuclear modification factor $R^p_{CP}$ for proton production
in \da collisions. The data are for 0-20\% to 60-90\% centralities
\cite{fm}. The solid line is the result of our calculation when all
contributions are taken into account, while the dashed line gives the
ratio when only the thermal contributions are
included.}
\end{figure}

The important properties of hadron production in  \da collisions
that we have learned from this study is that the protons are formed by
recombination at all $p_T$ and that the underlying partons that give
rise to their formation change smoothly from the soft component
  to the semi-hard shower
partons that are created by the hard partons.  The recombination
formalism allows us to calculate the  $p_T$ distribution in the
intermediate and higher $p_T$ regions with good agreement with the
data.  The contribution from the recombination of soft and shower
partons cannot be interpreted as a modification of the fragmentation
function, since the hard partons  in \da collisions are not
significantly affected by the cold medium that they traverse.
In this
treatment the Cronin effect of the proton spectrum is not induced by
transverse broadening due to initial scatterings, but is caused
mainly by the centrality dependence of the soft partons that
recombine.    $R^p_{CP}$ is higher than $R^{\pi}_{CP}$ because the
number of such recombining quarks is 3 instead of 2.  Our approach
cannot be viewed as being totally successful until the data for $p_T>
3$ GeV/c turn out to support our
predictions in the higher $p_T$ region.

We are grateful to R.\ J.\ Fries for providing us with the
parametrization of $f_i(k)$ that is given in \cite{hy} and used here
also.  Comments by V.\ Greco and C.\ M.\ Ko have been helpful.  This
work was supported, in part,  by the U.\ S.\ Department of Energy
under Grant No. DE-FG03-96ER40972  and by the Ministry of
Education of China under Grant No. 03113.


\begin{thebibliography}{99}

\bibitem{hy}
R.\ C.\ Hwa, and C.\ B.\ Yang, nucl-th/0403001.

\bibitem{jc} J.\ W.\ Cronin {\it et al.}, Phys.\ Rev.\ D {\bf 11},
3105 (1975).

\bibitem{fm} F.\ Matathias (PHENIX Collaboration), nucl-ex/0403029.

\bibitem{hy2}
R.\ C.\ Hwa, and C.\ B.\ Yang, hep-ph/0312271.

\bibitem{hy3}
   R.\ C.\ Hwa and C.\ B.\ Yang, nucl-th/0401001.

\bibitem{rh}
R.\ C.\ Hwa, Phys.\ Rev.\ D {\bf 22}, 1593 (1980).

\bibitem{hy4} R.\ C.\ Hwa, and C.\ B.\ Yang, Phys.\ Rev.\ C {\bf 67},
034902 (2003).

\bibitem{fm2}
F.\ Matathias (PHENIX Collaboration), talk given at Quark Matter
2004, Oakland, CA (2004).

\bibitem{gr}
V.\ Greco, C.\ M.\ Ko, and P.\ L\'{e}vai, Phys.\ Rev.\
Lett.\ {\bf 90}, 202302 (2003); Phys.\ Rev.\ C {\bf 68}, 034904 (2003).

\bibitem{fr}
R.\ J.\ Fries, B. M\"{u}ller, C.\ Nonaka and S.\ A.\
Bass, Phys.\ Rev.\ Lett.\ {\bf 90}, 202303 (2003);  Phys.\ Rev.\ C
{\bf 68}, 044902 (2003).

\bibitem{ssa}
S.\ S.\ Adler {\it et al.}, Phys.\ Rev.\ C {\bf 69},
034909 (2004).

\end{thebibliography}
\end{document}